# TIME-DEPENDENT ION TRANSPORT IN HETEROGENEOUS PERMSELECTIVE SYSTEMS

Yoav Green and Gilad Yossifon*

*Faculty of Mechanical Engineering, Micro- and Nanofluidics Laboratory, Technion–Israel Institute of Technology, Technion City 32000, Israel*

**ABSTRACT**

The current study extends previous analytical and numerical solutions of chronopotentiometric response of one-dimensional systems consisting of three layers to the more realistic two-dimensional heterogeneous ion-permselective medium. An analytical solution for the transient concentration-polarization problem, under the local electro-neutrality approximation and ideal permselectivity, was obtained using the Laplace transform and separation of variables. Then the two-dimensional electric potential was obtained numerically and was compared to the full Poisson-Nernst-Planck solution. It was then shown that the resultant voltage drop across the system varies between the initial Ohmic response and that of the steady-state accounting for concentration-polarization. Also, field-focusing effect in a two-dimensional system is shown to result in a faster depletion of ions at the permselective interface.

# I. INTRODUCTION

The passage of an electric current through a permselective medium (membranes or nanochannels) under an applied electric field is characterized by the formation of concentration gradients, which result in regions of depleted and enriched ionic concentration at opposite ends of the medium [1,2]. The formation of these concentration gradients and the resulting electric current are collectively termed concentration polarization (CP). Initially, in the low-voltage region, the *steady state* current-voltage (I-V) responds in an approximate Ohmic manner. At higher voltages, the current, which is diffusion-limited, eventually saturates at a limiting value when the ion concentrations are completely depleted at the permselective interface.

While the steady-state response of the system is of much importance and is used to characterize many attributes of the systems, such as the conductance in the Ohmic region[3–5], in fact it is the dynamical response which can provide much insight on the behavior of the system. From an experimental standpoint, it is crucial to know what is the characteristic time scale during which systems relax so that IV curves can be measured after minimizing time transient effects[6]. Theoretical works have investigated time transient effects in one-dimensional systems in numerous manners. As far back as 1901, the time-dependent concentration behavior of a solution at electrode interface was investigated[7–9]. This solution was then re-derived for a permselective system of a finite length [10]. Additional analysis for the concentration distribution in a semi-infinite domain, where at infinity the concentration has a bulk value and at the other end there exists a permselective interface has also been used in numerous works[11]. These solutions have been used to investigate the behavior of the concentration at the interface and the transition time until its complete depletion [12,13]. Also, chronopotentiometry experiments have investigated the time dependency of the electric potential [10,14,15] to a step-wise electric current. A number of works have investigated both the static [16,17] and dynamical[18–20] response of the extended space charge layer (SCL) which forms in the limiting current region.

However, until now, the effects of time-dependent CP phenomena in realistic heterogeneous three layered system (i.e., a permselective medium connected by two opposing microchambers – see Figure 1 for a schematic representation of such a system) has not been theoretically investigated in a thorough manner. In the current study we develop analytical and numerical models to describe time-dependent CP phenomenon for heterogeneous permselective systems. Where the term heterogeneity refers to the fact that the size of the permselective interface is smaller than the size of the system (see Figure 1 and references [21,22] for a complete discussion on heterogeneity in steady state and application related works [23,24]). In this work we shall focus on the electro-diffusive response of the system while neglecting electroconvection effects. In particular, the analytical solution is obtained under the assumption of local electroneutrality (LEN), hence, valid for Ohmic and limiting current conditions, while the numerical solution accounts also for the existence of the SCL. For simplicity we also neglect the surface charge on the microchamber walls and its associated surface-conduction effects [25,26].

In Sect. II we will define the theoretical model of and present its solution. In Sect. III we shall provide details regarding the numerical simulations. Whereas in Sect. IV we shall go into a lengthy discussion regarding the outcome of our solution. We then give concluding remarks in Sect. V. In attempt to keep the natural flow of this manuscript unhindered by lengthy mathematical derivations and yet at the same time avoid brevity coming at the expense of completeness, we provide full derivations in the Appendices.

## II. THEORETICAL MODEL

### A. Problem Definition

The equations governing the time dependent transport of a symmetric and binary $(z_+ = -z_- = 1)$ electrolyte of equal diffusivities $(D_+ = D_- = D)$ through a permselective medium are the dimensionless Poisson-Nernst-Planck (PNP) equations

$$c_{+,t} = \nabla \cdot [\nabla c_+ + c_+ \nabla \phi] = -\nabla \cdot \boldsymbol{j}_+ , \tag{1}$$

$$c_{-,t} = \nabla \cdot [\nabla c_- - c_- \nabla \phi] = -\nabla \cdot \boldsymbol{j}_- , \tag{2}$$

$$\nabla^2 \phi = -\frac{\rho_e}{2\varepsilon^2} , \tag{3}$$

wherein Eqs. (1) and (2) are the Nernst-Planck equations satisfying the continuity of ionic fluxes conditions. The cationic and anionic concentrations, $\tilde{c}_+$ and $\tilde{c}_-$, respectively, have been normalized by the bulk concentration $c_0$, where the tilde stands for the parameter in its dimensional form.. The spatial coordinates have been normalized by the diffusion length (DL) length $\tilde{L}$, the ionic fluxes have been normalized by $\tilde{D} c_0 / \tilde{L}$, while the time $\tilde{t}$ has been normalized by the diffusion time $\tilde{L}^2 / \tilde{D}$. Equation (3) is the Poisson equation for the electric potential, $\tilde{\phi}$, which has been normalized by the thermal potential $RT/F$ where $R$ is the universal gas constant, $T$ is the absolute temperature and $F$ is the Faraday constant. The charge density, $\rho_e$, appearing in Eq. (3) is normalized by $zFc_0$. The normalized Debye layer is $\varepsilon = \lambda_D / \tilde{L}$, with $\lambda_D = \sqrt{\varepsilon_0 \varepsilon_r RT / 2F^2 c_0}$ where $\varepsilon_0$ and $\varepsilon_r$ are the permittivity of vacuum and the relative permittivity of the electrolyte, respectively.

Under the LEN approximation [1,2,16,27], one can assume that $\varepsilon \ll 1$ (or alternately $\varepsilon^2 \nabla^2 \phi \sim 0$) within the microchambers, thus simplifying the equations. Hence, the Poisson equation Eq.(3) can be replaced with the approximation $c_+ = c_- = c$. Addition and subtraction of Eqs. (1) and (2) reduce to

$$c_t = \nabla^2 c,  \qquad (4)$$

$$\nabla \cdot (c \nabla \phi) = 0. \qquad (5)$$

While in the steady-state case the assumption of ideal permselectivity, $j_- = 0$, results in the expression for the electric potential (up to an additive constant)

$$\phi = \ln c, \qquad (6)$$

this is clearly not the case for non-steady transport as $j_- = 0$ holds only at the membrane interface and in a time-dependent problem can vary spatially within the microchambers (see Eq. (2)). It is noted that the problem of finding the concentration is decoupled from that of the electric potential, hence the concentration is solved first, through Eq.(4), and then the concentration solution is used for solving the electric potential, through Eq.(5).

### B. Geometry, boundary and initial conditions

Our model consists of a 3-layers system in which two microchambers are connected by a straight ideal cation permselective medium, wherein all three domains are of rectangular shape, as shown in Figure 1. The left microchamber, termed "region 1", is defined in the domain $x \in [0, L_1], y \in [0, H_1]$, the permselective medium termed "region 2" is defined in the domain $x \in [L_1, L_1 + d], y \in [0, h]$, while the right microchamber termed "region 3" is defined in the domain $x \in [L_1 + d, L_1 + d + L_3], y \in [0, H_3]$. Such a geometry realistically describes systems that have been the subject of numerous recent experimental and numerical works [3,4,28–37]. Additionally, this geometry can also describe a periodic array of permselective regions (e.g. nanochannel array/heterogeneous membrane) in the *y* direction. The spatial coordinates have been normalized by the DL length, $\tilde{L}$ ( $\tilde{L}$ can be chosen arbitrarily as either $\tilde{L}_1$ or $\tilde{L}_3$ [5]). Without loss of generality, we shall formulate the solution for general values of the dimensionless $L_1$ and $L_3$ while we shall remember that at least one of these values when normalized is unity.

Assuming fixed volumetric charge density, $N$, accounting for the (negative) surface charge within the nanoslot, as in classical models of permselective membranes[36,38], the space charge within all three regions $(n=1,2,3)$ can be written as follows

$$\rho_{e,n} = c_+ - c_- - N\delta_{n,2} \ , \tag{7}$$

where $\delta_{n,2}$ is Kronecker's delta. The LEN approximation in the microchambers and cross-sectional electro-neutrality within the permselective medium corresponds to $\rho_{e,n} \approx 0$. The case of $N \gg 1$ approximates the conditions of an ideal permselective membrane/nanochannel, requiring that $c_+ \approx N$ and $c_- \approx 0$.

The boundary conditions (BCs) and initial conditions (ICs) are

$$c(x=0,y,t) = c(x=L_1+d+L_3,y,t) = 1 \ , \tag{8}$$

$$c_y(x,y=0,t) = c_y(x,y=H_i,t) = 0, \quad i=1,3 \ , \tag{9}$$

$$c_x(x=L_1,y,t) = \begin{cases} -i/2 & 0 \le y \le h \\ 0 & \text{else} \end{cases} , \tag{10}$$

$$c_x(x=L_1+d,y,t) = \begin{cases} -i/2 & 0 \le y \le h \\ 0 & else \end{cases} , \tag{11}$$

$$c_{1,3}(x,y,t=0) = 1 \ , \tag{12}$$

where Eq. (8) stands for the stirred bulk electrolyte concentration at the opposite microchannel entrances. Equation (9) is the constraint of no-penetration of ions ($\boldsymbol{j}_\pm \cdot \boldsymbol{n} = 0$, where $n$ is the coordinate normal to the surface at the solid walls or symmetry planes). Equation (9) is used in conjunction with requiring electrical insulation ($\partial \phi / \partial n = 0$), at the walls or symmetry planes. Equations (10) and (11) are the simplifying assumption of a uniform ionic current density right at the ideal cationic perm-selective interface (i.e. $\boldsymbol{j}_- \cdot \boldsymbol{n} = 0$) [21], with $i$ ($= |\boldsymbol{i}|$) being the *uniform* dimensionless current density. In an ideal permselective medium $\tilde{\boldsymbol{i}} = F\tilde{\boldsymbol{j}}_+$, or in dimensionless form $\boldsymbol{i} = \boldsymbol{j}_+$, the current density has been normalized by $F\tilde{D}c_0 / \tilde{L}$. Equation (12) is the IC of a uniform bulk concentration within the microchambers under an equilibrium condition. Note that since we are assuming an ideal permselective medium the counterion concentration within region 2 is fixed in time $c_+ \approx N$ and $c_- \approx 0$.

### C. Concentration solution

Using the Laplace Transform, Eq. (4) transforms into the Helmholtz equation. This equation is then solved using a separation of variables technique as described in Appendix A. The solution for the concentration in each region are given by the following expressions

$$c_1(x,y,t) = 1 - \frac{I}{2H_1}x - \frac{I}{hH_1}\sum_{n=1}^{\infty}\frac{\sin\left(\lambda_n^{(1)}h\right)\cos\left(\lambda_n^{(1)}y\right)\sinh\left(\lambda_n^{(1)}x\right)}{\left(\lambda_n^{(1)}\right)^2 \cosh\left(\lambda_n^{(1)}L_1\right)} \\ -\frac{I}{H_1L_1}\sum_{m=1}^{\infty}(-1)^m\frac{\sin\left(\gamma_m^{(1)}x\right)}{\left(\gamma_m^{(1)}\right)^2}e^{-\left(\gamma_m^{(1)}\right)^2 t} - \frac{2I}{hH_1L_1}\sum_{n,m=1}^{\infty}(-1)^m\frac{\sin\left(\lambda_n^{(1)}h\right)\cos\left(\lambda_n^{(1)}y\right)\sin\left(\gamma_m^{(1)}x\right)e^{-\left(\kappa_{mn}^{(1)}\right)^2 t}}{\lambda_n^{(1)}\left(\kappa_{mn}^{(1)}\right)^2}, \quad (13)$$

$$c_{+,2}(x,y,t) = N, \quad c_{-,2}(x,y,t) = 0 , \tag{14}$$

$$c_3(x,y,t) = 1 + \frac{I}{2H_3}(L_1 + d + L_3 - x) + \frac{I}{hH_3}\sum_{n=1}^{\infty}\frac{\sin(\lambda_n^{(3)}h)\cos(\lambda_n^{(3)}y)\sinh[\lambda_n^{(3)}(L_1+d+L_3-x)]}{(\lambda_n^{(3)})^2 \cosh(\lambda_n^{(3)}L_3)}$$
$$+ \frac{I}{H_3 L_3}\sum_{m=1}^{\infty}(-1)^m \frac{\sin[\gamma_m^{(3)}(L_1+d+L_3-x)]}{(\gamma_m^{(3)})^2}e^{-(\gamma_m^{(3)})^2 t}$$
$$+ \frac{2I}{hH_3 L_3}\sum_{n,m=1}^{\infty}(-1)^m \frac{\sin(\lambda_n^{(3)}h)\cos(\lambda_n^{(3)}y)\sin[\gamma_m^{(3)}(L_1+d+L_3-x)]e^{-(\kappa_{nm}^{(3)})^2 t}}{\lambda_n^{(3)}(\kappa_{mn}^{(3)})^2}$$
$$\tag{15}$$

with the current (normalized by $F\tilde{D}c_0$) and eigenvalues defined as

$$I = ih , \tag{16}$$

$$\lambda_n^{(k)} = \frac{\pi n}{H_k}, \gamma_m^{(k)} = \frac{\pi(2m-1)}{2L_k}, \kappa_{mn}^{(k)} = \sqrt{(\lambda_n^{(k)})^2 + (\gamma_m^{(k)})^2} , \tag{17}$$

and the superscript $k = 1,3$ defines the region. The first three terms in Eqs. (13) and (15) are the steady state solution previously derived in Refs. [5,21,22,39]. The fourth and fifth terms are the time transient solutions corresponding to the one-dimensional transient solution[10] and the two-dimensional field focusing transient, respectively. In the permselective area the co-ion and counterion concentrations are held constant. The concentrations at the interface, at $y = 0$, are

$$c_1(x = L_1, y = 0, t) = 1 - \frac{IL_1}{2H_1} - I\bar{f}_1 - I\bar{g}_1(t) , \tag{18}$$

$$c_3(x = L_1 + d, y = 0, t) = 1 + \frac{IL_3}{2H_3} + I\bar{f}_3 + I\bar{g}_3(t) , \tag{19}$$

$$\bar{f}_k = \frac{1}{hH_k} \sum_{n=1}^{\infty} \frac{\sin\left(\lambda_n^{(k)}h\right)\tanh\left(\lambda_n^{(k)}L_k\right)}{\left(\lambda_n^{(k)}\right)^2} , \qquad (20)$$

$$\bar{g}_k(t) = \frac{1}{H_k L_k} \sum_{m=1}^{\infty} \frac{\sin\left(\gamma_m^{(k)}L_k\right)}{(-1)^m \left(\gamma_m^{(k)}\right)^2} e^{-\left(\gamma_m^{(k)}\right)^2 t} + \frac{2}{hH_k L_k} \sum_{n,m=1}^{\infty} \frac{\sin\left(\lambda_n^{(k)}h\right)\sin\left(\gamma_m^{(k)}L_k\right)e^{-\left(\kappa_{mn}^{(k)}\right)^2 t}}{(-1)^m \lambda_n^{(k)}\left(\kappa_{mn}^{(k)}\right)^2} , \qquad (21)$$

where $\bar{f}_k = \bar{f}_k(L_k, H_k, h)$ represents the steady state contribution of the field focusing to the solution and is a function of the geometry in each region. The behavior of the $\bar{f}_k$ functions in the varying limits of heterogeneity, i.e. large and small $h/H_k$, has been recently investigated [22]. Whereas $\bar{g}_k(t) = \bar{g}_k(t; L_k, H_k, h)$ represents the time transient contribution with the first term representing the one-dimensional decay and the second term representing the two-dimensional decay.

### D. Electric potential solution

As previously mentioned, to find the time-dependent electric potential, one must solve Eq.(5), $\nabla \cdot (c\nabla\phi) = 0$. This equation is not explicitly time dependent but rather implicitly through the concentration. A semi-analytical solution for the one-dimensional potential drop case is given in Ref. [10] and is rederived in Appendix C to include asymmetric microchannel lengths. In general, Eq. (5) in the two-dimensional case cannot be solved analytically, except at the two extreme cases of cases of $t=0$ and $t\to\infty$, but rather requires numerical evaluation. The appropriate BCs are (Figure 1)

$$\phi_x(x=0, y, t) = -I/2H_1, \quad \phi(x = L_1 + d + L_3, y, t) = 0 , \qquad (22)$$

$$\phi_y(x, y=0, t) = \phi_y(x, y=H_i, t) = 0, \quad i = 1, 3 , \qquad (23)$$

$$\phi_x(x = L_1, y, t) = \begin{cases} -i/2c & 0 \leq y \leq h \\ 0 & \text{else} \end{cases}, \tag{24}$$

$$\phi_x(x = L_1 + d, y, t) = \begin{cases} -i/2c & 0 \leq y \leq h \\ 0 & \text{else} \end{cases}, \tag{25}$$

$$\phi_{1,3}(x, y, t = 0) = 0, \tag{26}$$

$$\phi_2(x, y, t = 0) = -\ln N, \tag{27}$$

where Eq. (22) is the condition of a grounded side and driving electrical current. Equation (23) is the electrical insulation BC. Equations (24) and (25) are, once more, the simplifying assumption of a uniform ionic current density at the permselective interfaces. Equations (26) and (27) are the IC of a system at equilibrium with the latter being the Donnan potential within the permselective medium. To facilitate the Donnan potential jump we require the continuity of the electrochemical potential of the counterions

$$\mu_+(x, y, t) = \mu(x, y, t) = \ln c_+(x, y, t) + \phi(x, y, t), \tag{28}$$

at the microchannel-permselective medium interface

$$\mu_1(x = L_1, y, t) = \mu_2(x = L_1, y, t), \tag{29}$$

$$\mu_2(x = L_1 + d, y, t) = \mu_3(x = L_1 + d, y, t). \tag{30}$$

Solving this problem will yield the desired solution $\phi(x = 0, y, t) = V(t)$. We note here that the co-ion electrochemical potential of

$$\mu_-(x, y, t) = \ln c_-(x, y, t) - \phi(x, y, t), \tag{31}$$

cannot be used, because in region 2 due to the assumption of ideal permselectivity, $c_- \approx 0$. However, prior to presenting and discussing the numerically computed solution, it is beneficial to discuss the two attainable exact solutions ($t=0$ and $t \to \infty$). We shall start off with the steady state solution which was solved in our previous work[5]. For the sake of brevity we shall only give a brief outline of the derivation of the solution. Within an ideal permselective medium the counterions concentration is constant (Eq.(14)) and under the assumption that the top and bottom surfaces are insulating, we have from the ion-flux continuity equation (Eq. (1))

$$i = -N\phi_x \Rightarrow \phi_2 = -\frac{I}{Nh}x + \bar{\phi}_2 \ . \tag{32}$$

In general, $\bar{\phi}_2$ is not a constant but is rather time dependent $\left(\bar{\phi}_2 = \bar{\phi}_2(t)\right)$. In steady state $(t \to \infty)$, where it can be assumed that $j_- = 0$ everywhere and not just within the permselective medium, the potential in the microchambers can be solved from Eq.(2)

$$\phi_1 = \ln c_1 + V, \qquad \phi_3 = \ln c_3 \ , \tag{33}$$

The solution for $\bar{\phi}_2$ and an $I-V$ relation can be solved by using the electrochemical continuity requirement given in Eqs.(29)-(30) by requiring continuity at the point $y=0$ rather than at the entire cross-section [5]

$$\bar{\phi}_2 = \frac{I}{hN}(L_1 + d) - \ln N + 2\ln\left(1 + \frac{IL_3}{2H_3} + I\bar{f}_3\right), \tag{34}$$

$$V_s = \frac{Id}{hN} + 2\ln\left[\frac{1 + \frac{IL_3}{2H_3} + I\bar{f}_3}{1 - \frac{IL_1}{2H_1} - I\bar{f}_1}\right], \tag{35}$$

where the subscript $s$ stands for steady-state conditions. In the Ohmic region, for the case of small currents $(I \ll 1)$, Eq. (35) is expanded to give the overall steady-state conductance per unit width (normalized by $\tilde{D}F^2 c_0/RT$) of the 3-layers system

$$\sigma_s = \frac{1}{R_s} = \frac{I}{V_s} = \left( \frac{d}{hN} + \frac{L_1}{H_1} + \frac{L_3}{H_3} + 2\bar{f}_1 + 2\bar{f}_3 \right)^{-1}, \tag{36}$$

and $R_s$ is the system's Ohmic resistance.

At time $t=0$, prior to application of an electric current, the concentration in each of the regions, excluding the electric-double layers (EDLs) interfacing the microchannel-permselective medium interface (LEN approximation), is spatially independent $(c_{1,3}=1, c_{+,2}=N, c_{-,2}=0)$. Hence the governing equation for the electric potential is reduced from $\nabla \cdot (c \nabla \phi) = 0$ to the Laplace equation

$$\nabla^2 \phi = 0 \ . \tag{37}$$

In Appendix B we derive the solution for the electric potential distribution in all the regions at $t=0$. The solution differs from the one given in Eqs. (32)-(33). We also derive the initial/rest conductance of the electrolyte as a function of the geometry

$$\sigma_0 = \frac{1}{R_0} = \frac{I}{V_0} = \left( \frac{d}{hN} + \frac{L_1}{2H_1} + \frac{L_3}{2H_3} + \bar{f}_1 + \bar{f}_3 \right)^{-1} . \tag{38}$$

where the subscript 0 stands for initial time $(t=0)$ conditions. We note that both Eqs. (36) and (38) are a property of the geometry and permselective counterion concentration. For the sake of generality, following the derivation for $t=0$ given in Appendix B of this work and in combination for the 3D potential distribution given in the Appendix of our previous work [5], we provide the initial conductance response given by Eq. (38) for a three-dimensional geometry with

$$\sigma_0 = \frac{1}{R_0} = \frac{I}{V_0} = \left( \frac{d}{Nwh} + \frac{L_1}{2W_1 H_1} + \frac{L_3}{2W_3 H_3} + \overline{f}_1 + \overline{f}_3 \right)^{-1} \quad (39)$$

with $w$ being the width of the permselective interface and $W_{1,3}$ the width of the according regions (see Fig 1. of Ref. [5] for a schematic). The full 3D expression for the $\overline{f}$ functions are given by Eq. 26 of Ref. [5].

### III. NUMERICAL SIMULATIONS

To verify our results we solved both the LEN approximation (marked by $\varepsilon = 0$ numerical) for the electric potential and the fully coupled PNP equations (marked by $\varepsilon = 10^{-4}$ numerical) given by Eqs.(1)-(3) using the finite elements program Comsol$^{TM}$ for the two-dimensional geometry described in Figure 1. The LEN model was solved using the Partial Differential Equation module, while the PNP equation were solved using the Transport of Diluted Species and Electrostatic modules in Comsol.

It can be observed from Figure 1 that, based on the BCs, for the LEN model region 3 can be solved independently of the remaining regions. After which, region 2 can be solved by requiring continuity (with region 3) of the electrochemical potential at the entire interface as given by Eq.(30). Thereafter, using Eq. (29) region 1 is solved. Finally the potential at $\phi(x=0, y, t) = V(t)$ is evaluated. The LEN numerical model was solved using the BCs specified in Sections. II.B and II.D. In contrast, the PNP does not require internal BCs at the interfaces between two neighboring regions (such as Eqs. (29)-(30)) as the continuity of ionic fluxes and electric fluxes is accounted for by Comsol.

Unlike the LEN model $(\varepsilon = 0)$, the PNP model $(\varepsilon \neq 0)$ accounts for both non-electroneutral effects (i.e. EDLs and emergence of SCL) and non-ideal membrane permselectivity. The IC ($t = 0$) for the PNP simulations, which include the contribution of the EDL, was calculated by applying a zero current, $I = 0$. This equilibrated solution was also used as an initial guess in the current-voltage sweep simulation. Both results will be shown in the following section. We wish to point out that the LEN simulations run substantially quicker (2-3 orders of magnitude) than the PNP simulations. This is due to the need to mesh the EDL at the interface in an extremely fine manner which is absent in the LEN simulation. At the two permselective interfaces we use a minimal mesh triangular mesh element of $\varepsilon / 30$ with $\varepsilon = 10^{-4}$.

## IV. RESULTS AND DISCUSSION

### A. Time evolution of concentration-polarization

In Figure 2 the time evolution of the concentration profiles for the depleted region are shown for a one-dimensional and two-dimensional case at the limiting current (corresponding to vanishing concentration at the permselective interface in region 1). It is observed that the two-dimensional system reaches its steady-state interfacial value faster than the one-dimensional system. This can be explained due the following scaling argument of the diffusion equation (Eq. (4)). The left hand side scales as $c_0 / \tilde{t}$ while the right hand side scales as $n_d c_0 / \tilde{L}^2$ with $n_d$ being the dimensionality of the system. This gives a characteristic time $\tilde{t} \approx \tilde{L}^2 / (n_d \tilde{D})$ (in dimensional form) that decreases with increasing $n_d$. This is also verified in Figure 3 depicting the time required for the concentration at the interface ($x = L_1$) to reach a quasi steady-state value of $1.001 c_s$ (or accordingly from Eq.(18), when $|\overline{Ig}_1(t)| = 0.001 c_s$). It is shown that as a system diverges from a homogenous one-dimensional system $(h = H_{1,3})$ into a heterogeneous two-dimensional system the typical time to depletion of ions at the interface decreases. Additionally, we show that this time increases as the current is increased. This is an expected result as for larger currents the degree of depletion increases and more time is required for ions to be removed from the interface. From an intuitive standpoint, the two-dimensional case reaches steady-state quicker as the transition is from a uniform concentration of unity to a sharp logarithmic profile, versus the one-dimensional case where the linear profile is achieved.

### B. Time evolution of the electric potential

In Figure 4 the electric potential as function of time is plotted in all three regions for a two-dimensional model. The numerically calculated potentials overlaps with the analytical solutions for $t = 0$ and $t \to \infty$. Interestingly, it is observed that while the potential drop across the anodic and cathodic microchambers is initially similar for the specified geometry. Also observed, the cathodic and the permselective media potential drops do not change substantially over time whereas in contrast, the potential drop within the anodic microchannel (and its associated Donnan jump) increases over time. As a result the total voltage drop across the system $V(t)$ also changes substantially. This corresponds to an increase in the resistance from $t = 0$ (Eq.(38)) to that at steady-state $t \to \infty$ (Eq.(36)). In actuality, it should be stated that the initial resistance given by Eq.(38) is independent on the current regime (Ohmic or Limiting or Over-Limiting). While Eq.(36) provides a simple expression for the steady-state resistance in the Ohmic region, for larger currents the voltage is non-linearly dependent on the current and must be evaluated from Eq. (35). Another interesting feature regarding Eqs. (36) and (38) is that the ratio between the steady-state and initial conductance approaches 2 when the permselective medium's resistance, $d/hN$, vanishes due to the inverse dependence on the fixed volumetric charge $N$ where in the ideal permselective case $N \gg 1$.

Figure 5 depicts the time evolution of the voltage drop $V(t)$ for both the one-dimensional and two-dimensional cases.. The one-dimensional case, exhibits an excellent agreement between the LEN simulation $(\varepsilon = 0)$ and theoretical prediction for the initial and steady-state cases. For the one-dimensional model, there exists a simpler way to numerically evaluate the potential drop over the entire system $V(t)$. In Appendix C we provide a derivation for semi-analytical expression that needs to be evaluated numerically. The advantage of such a method is that one does not need to resort to numerical simulations. However, one-dimensional simulations are important stepping to developing two-dimensional LEN simulations. We have not added this curve in Figure 3a as it completely overlaps the simulated curve given by the solid magenta line.

We shall now explain the observed deviations of the two-dimensional voltage drop (Figure 5b) as well as the deviation due the loss of LEN (PNP model). In the two-dimensional case the numerical solution under-predicts the theoretical prediction for the steady-state voltage(Figure 5b). This difference stems from the latter satisfying the electrochemical continuity at a single point at each of the permselective interfaces, while in the simulations, the former satisfy the continuity across the entire permselective medium interface. For the case of solving the fully coupled PNP $(\varepsilon = 10^{-4})$ equations it can be observed that the total voltage drop has a larger deviation from the analytically predicted steady-state result. A similar deviation is also visible in the I-V curves depicted in Figure 6 (at steady-state conditions). For the one-dimensional case (Figure 6a) the analytical solution,

Eq.(35), and the numerical LEN solution are identical and show the expected saturation of the current to the limiting value corresponding to an infinite resistance[1]. At high voltages the PNP solution exhibits a large but finite resistance corresponding to the creation of the SCL[16,17]. This space charge increases the overall conductance of the system relative to that of the LEN conductance. In a complimentary manner this results in a decrease of the resistance of the $\varepsilon = 10^{-4}$ relative to the $\varepsilon = 0$ case. This is also true for the two-dimensional case presented in Figure 6b. This also explains the additional decrease of the steady-state resistance depicted in Figure 5 for the cases of $\varepsilon = 10^{-4}$.

## V. CONCLUSIONS

We derived an analytical solution, using the Laplace transform and separation of variable technique, for the temporal-spatial concentration distribution in a two-dimensional three layered system (Eqs.((13)-(15))). We then proceed to investigate the time dependent behavior exhibited by the system. In steady state the concentration profiles for a one-dimensional and two-dimensional system are linear and logarithmic-like, respectively. Due to the field focusing effect in a two-dimensional system a faster depletion of ions occurs at the permselective interface corresponding to a decreased characteristic time with increasing dimensionality. We then derived an expression for the initial resistance of the system as function of the geometry and permselective fixed volumetric charge concentration. In the case where the permselective medium resistance $d/hN$ is substantially lower than the remaining resistors, the ratio between the Ohmic steady-state and initial resistors is $R_s / R_0 = 2$. This is explained due to the fact that in steady-state the current is transported only by the positively charged counterions, due to ideal permselectivity of the membrane, where at initial times both the counterions and co-ions contribute equally to the current. In the limit of vanishing permselectivity, the concentration is constant and uniform throughout the system resulting in the $t=0$ Ohmic response being correct at all times.

The electric potential was then calculated numerically under the LEN approximation as well as solving the fully-coupled PNP equations. The two different sets of numerical simulations confirm that the resultant electric potential evolves between the two analytical solutions for $t=0$ and $t\to\infty$. In particular, we obtained an almost abrupt jump in the potential for the two-dimensional case which is associated with the fast interfacial depletion at the anodic side of the permselective medium. As time evolves the Donnan potentials drops across the anodic (cathodic) side of the membrane is increasing (decreasing) as expected from Eqs. (29) and (30), (i.e. $\Delta\phi_{Donnan} = \ln(N/c_+)$), since the interfacial concentration within the microchamber $c_+$ is depleted (enriched) with time. In contrast the voltage drop across the permselective medium is constant due to its Ohmic behavior, i.e. constant resistance. It should be pointed out that in two-dimensional the numerically calculated resultant voltage is lower than the predicted steady-state response. This is due to the fact that the steady-state theoretical model assumes continuity of the electric potential only at a single point of the permselective interface whereas the numerical model ensures continuity at the interface. The difference is best observed in the I-V curves which show that a minute difference exists solely in the limiting region.

As suggested in the current study, the measured initial voltage drop of a chronopotentiometric experiment provides, through Eq. (38), means to extract the effective fixed volumetric charge of the permselective medium, $\tilde{N}$. This is advantageous over the steady-state Ohmic current-voltage response (Eq.(36)) which necessitates slow sweep rates (in relation to the diffusion time scale) of the voltage in order to obtain a quasi steady-state response.

**ACKNOWLEDGMENTS**

We thank Dr. Sinwook Park and Dr. Ory Schnitzer for fruitful discussions and helpful comments. This work was supported by ISF Grant No. 1078/10. We thank the Technion RBNI (Russell Berrie Nanotechnology Institute) and the Technion GWRI (Grand Water Research Institute) for their financial support.

# APPENDIX A: CONCENTRATION SOLUTION DERIVATION

Using the Laplace Transform for the time coordinate

$$C(x, y, s) = \int_0^\infty c(x, y, t) e^{-st} dt \tag{40}$$

on Eq.(4) gives

$$sC - c_{t=0} = C_{xx} + C_{yy}, \tag{41}$$

which is the inhomogeneous Helmholtz equation and $c_{t=0}$ is the initial condition in Eq.(12). The modified BCs for region 1 are

$$C(x = 0, y, s) = 1/s, \tag{42}$$

$$C_y(x, y = 0, s) = C_y(x, y = H_1, s) = 0, \tag{43}$$

$$C_x(x = L_1, y, s) = \begin{cases} -i/(2s), & 0 \le y \le h \\ 0, & \text{else} \end{cases}. \tag{44}$$

A separation of variables method

$$C_h(x, y, s) = X(x, s) Y(y, s), \tag{45}$$

is used to solve the homogenous equation in Eq.(41). It is clear from Eq. (45) that all functions are explicitly dependent on $s$, yet for the sake of brevity, from this point on we shall write this implicitly. Inserting Eq. (45) in Eq. (41) gives

$$s - \frac{X''(x)}{X(x)} = \frac{Y''(y)}{Y(y)} = -\lambda^2. \tag{46}$$

The solution of Eq. (46) is

$$Y = A\cos(\lambda y) + B\sin(\lambda y). \tag{47}$$

Use of BC Eq. (43) gives

$$B = 0, \quad \lambda_n = \pi n / H_1, \quad n = 1, 2, 3... \tag{48}$$

So that

$$Y_n(y) = A_n \cos(\lambda_n y) . \tag{49}$$

For the case $\lambda^2 = n = 0$, it is easy to see that the $Y_{n=0} = A_0$ and the governing equation and it's solution are

$$sC_0 - c_{t=0} = C_{0xx} , \tag{50}$$

$$C_0 = E_0 \sinh(\sqrt{s}x) + F_0 \cosh(\sqrt{s}x) + \frac{1}{s} , \tag{51}$$

where we have added the inhomogeneous component $c_{t=0}$ which gives the $1/s$ term. Then from Eq.(46) and Eq.(48) one obtains

$$X_n(x) = E_n \sinh\left(\sqrt{\lambda_n^2 + s}\,x\right) + F_n \cosh\left(\sqrt{\lambda_n^2 + s}\,x\right) . \tag{52}$$

Hence the complete solution is

$$C(x,y,s) = \frac{1}{s} + E_0 \sinh(\sqrt{s}x) + F_0 \cosh(\sqrt{s}x) + \sum_{n=1}^{\infty}\left[E_n \sinh\left(\sqrt{\lambda_n^2 + s}\,x\right) + F_n \cosh\left(\sqrt{\lambda_n^2 + s}\,x\right)\right]\cos\lambda_n y . \tag{53}$$

Use of BC Eq. (42) requires that

$$F_0 = 0, F_n = 0 . \tag{54}$$

For the BC of current flux density conservation (Eq.(44)), we then take the $x$ derivative of Eq. (53) and find the Fourier coefficients

$$E_0 = -\frac{I}{2H_1 s\sqrt{s}\cosh(\sqrt{s}L_1)}, E_n = -\frac{I\sin(\lambda_n h)}{\lambda_n H_1 hs\sqrt{\lambda_n^2 + s}\cosh\left(\sqrt{\lambda_n^2 + s}L_1\right)} , \tag{55}$$

with $I = ih$. Inserting Eqs. (54) and (55) into Eq.(53) gives

$$C(x,y,s) = \frac{1}{s} - \frac{I\sinh(\sqrt{s}x)}{2H_1 s\sqrt{s}\cosh(\sqrt{s}L_1)} - \frac{I}{H_1 h}\sum_{n=1}^{\infty}\left[\frac{\sin(\lambda_n h)\sinh\left(\sqrt{\lambda_n^2 + s}\,x\right)\cos(\lambda_n y)}{\lambda_n s\sqrt{\lambda_n^2 + s}\cosh\left(\sqrt{\lambda_n^2 + s}L_1\right)}\right] . \tag{56}$$

We calculate the Inverse Laplace transform using the residue theorem

$$c(x,y,t) = \frac{1}{2\pi j}\lim_{T\to\infty}\int_{\alpha-jT}^{\alpha+jT} C(x,y,s)e^{st}ds = \text{Res}\left[C(x,y,s)e^{st}, s_l\right], \quad (57)$$

for all $t>0$, where $s_l$ is a pole of $C(x,y,s)$ and $j$ is the imaginary unit. The first term in Eq. (56) has a single pole at $s=0$ yielding

$$c_I(x,y,t) = 1. \quad (58)$$

The second term has one pole at $s=0$ and an infinite series of poles

$$s_m = -\frac{\pi^2(2m-1)^2}{4L_1^2} = -\gamma_m^2, \quad m=1,2,3... \quad (59)$$

Thus, yielding from the second term in Eq. (56)

$$c_{II}(x,y,t) = -\frac{I}{2H_1}x - \frac{I}{H_1 L_1}\sum_{m=1}^{\infty}\frac{\sin(\gamma_m x)}{\gamma_m^2 \cos(\pi m)}e^{-\gamma_m^2 t}. \quad (60)$$

The term $\cos\pi m$ is simply $(-1)^m$. Finally the third term has poles at $s=0$ and

$$s_{mn} = -\gamma_m^2 - \lambda_n^2 = -\kappa_{mn}^2 \quad m=1,2,3... \quad n=1,2,3... \quad (61)$$

transforming the third term of Eq. (56). Into

$$c_{III}(x,y,t) = -\frac{I}{H_1 h}\sum_{n=1}^{\infty}\frac{\sin(\lambda_n h)\cos(\lambda_n y)\sinh(\lambda_n x)}{\lambda_n^2 \cosh(\lambda_n L_1)} - \frac{2I}{L_1 H_1 h}\sum_{n,m=1}^{\infty}\frac{\sin(\lambda_n h)\cos(\lambda_n y)\sin(\gamma_m x)e^{-\kappa_{mn}^2 t}}{\lambda_n(\gamma_m^2 + \lambda_n^2)\cos(\pi m)}$$

$$(62)$$

Then the solution for the concentration in region 1 is the sum of Eqs.(58), (60) and (62)

$$c_1(x,y,t) = 1 - \frac{I}{2H_1}x - \frac{I}{hH_1}\sum_{n=1}^{\infty}\frac{\sin(\lambda_n^{(1)}h)\cos(\lambda_n^{(1)}y)\sinh(\lambda_n^{(1)}x)}{(\lambda_n^{(1)})^2\cosh(\lambda_n^{(1)}L_1)}$$

$$-\frac{I}{H_1 L_1}\sum_{m=1}^{\infty}(-1)^m\frac{\sin(\gamma_m^{(1)}x)}{(\gamma_m^{(1)})^2}e^{-(\gamma_m^{(1)})^2 t} - \frac{2I}{hH_1 L_1}\sum_{n,m=1}^{\infty}(-1)^m\frac{\sin(\lambda_n^{(1)}h)\cos(\lambda_n^{(1)}y)\sin(\gamma_m^{(1)}x)e^{-(\kappa_{nm}^{(1)})^2 t}}{\lambda_n^{(1)}(\kappa_{nm}^{(1)})^2}$$

$$(63)$$

We note that for the case $h = H_1$, the two-dimensional transient solution reverts to the one-dimensional solution given in Ref. [10]. Similarly to the above shown procedure one can find the solution for region 3 using the relevant BCs.

## APPENDIX B: INITIAL ELECTRIC POTENTIAL SOLUTION

At $t = 0$ when the concentration in each of the regions is uniform the equation governing the electric potential within the microchambers, outside EDLs at the microchannel-permselective medium interface, is simply the Laplace equation

$$\nabla^2 \phi = 0 \tag{64}$$

Given that the BCs, previously given in Sect. II.D, for the electric potential are similar to the BCs of the concentration given in Sect II.B, this suggests that the electric potential solution has a similar form to that of the concentration (as was previously shown in Ref.[5] which solved a similar problem governed by the Laplace equation, albeit a different physical situation). It can be shown that the electric potential in region 3 is

$$\phi_3(x,y) = \frac{I}{2H_3}(L_1 + d + L_3 - x) + \frac{I}{hH_3} \sum_{n=1}^{\infty} \frac{\sin \lambda_n^{(3)} h \sinh\left[\lambda_n^{(3)}(L_1 + d + L_3 - x)\right] \cos \lambda_n^{(3)} y}{\left(\lambda_n^{(3)}\right)^2 \cosh\left(\lambda_n^{(3)} L_3\right)} \tag{65}$$

The electric potential in region 2 remains unchanged

$$\phi_2(x,y) = -\frac{I}{hN}x + \bar{\phi}_{2,0} \tag{66}$$

and requiring continuity of electrochemical potential (Eq.(30)) gives

$$\bar{\phi}_{2,0} = \frac{I}{hN}(L_1 + d) - \ln N + \frac{IL_3}{2H_3} + I\bar{f}_3 . \tag{67}$$

The potential in region 1 is accordingly

$$\phi_1(x,y,t) = V_0 - \frac{I}{2H_1}x - \frac{I}{hH_1}\sum_{n=1}^{\infty}\frac{\sin(\lambda_n^{(1)}h)\sinh(\lambda_n^{(1)}x)\cos(\lambda_n^{(1)}y)}{(\lambda_n^{(1)})^2 \cosh(\lambda_n^{(1)}L_1)}. \tag{68}$$

The initial potential jump, $V_0$, at time $t = 0$ can be found using Eq.(29) yielding

$$V_0 = I\left(\frac{d}{Nh} + \frac{L_1}{2H_1} + \frac{L_3}{2H_3} + \overline{f_1} + \overline{f_3}\right) \tag{69}$$

which corresponds to an initial conductance of

$$\sigma_0 = \frac{I}{V_0} = \left(\frac{d}{Nh} + \frac{L_1}{2H_1} + \frac{L_3}{2H_3} + \overline{f_1} + \overline{f_3}\right)^{-1}. \tag{70}$$

## APPENDIX C: ONE-DIMENSIONAL TIME DEPENDENT VOLTAGE DERIVATION

In this appendix we shall derive an expression for the potential drop over the entire system for the one-dimensional case. A solution for the potential drop over the entire system for the one-dimensional case was previously presented in Ref. [10]. For completeness reasons we re-derive the expression and extend it to the case of non-symmetric microchamber lengths which also includes an expression for the permselective medium resistance. Following Eq.(5), the potential drop over region 3 is given by the integral

$$\Delta\phi_3(t) = -\int_{L_1+d}^{L_1+d+L_3} \frac{i}{2c_3(x,t)}dx. \tag{71}$$

The potential drop over the permselective medium remains unchanged during the transition phase

$$\Delta\phi_2 = -i\frac{d}{N}, \tag{72}$$

while, the Donnan potential drops at the opposite interface are obtained, using Eqs.(29)-(30), as

$$\Delta\phi_{D23} = \ln\left[\frac{N}{c_3(x = L_1 + d)}\right], \tag{73}$$

$$\Delta\phi_{D12} = \ln\left[\frac{c_1(x = L_1)}{N}\right]. \tag{74}$$

Leading to a total Donnan potential drop *difference* of

$$\Delta \phi_D = \ln \left[ \frac{c_1(x = L_1)}{c_3(x = L_1 + d)} \right]. \tag{75}$$

Similarly, the potential drop in region 1 is

$$\Delta \phi_1(t) = -\int_0^{L_1} \frac{i}{2c_1(x,t)} dx. \tag{76}$$

Thus the voltage drop over the entire system is simply

$$\Delta \phi(t) = \Delta \phi_1 + \Delta \phi_2 + \Delta \phi_3 + \Delta \phi_D, \tag{77}$$

which can be written specifically

$$V(t) = -\Delta \phi(t) = \int_0^{L_1} \frac{i}{2c_1(x,t)} dx + i \frac{d}{N} + \int_{L_1+d}^{L_1+d+L_3} \frac{i}{2c_3(x,t)} dx - \ln \left[ \frac{c_1(x = L_1)}{c_3(x = L_1 + d)} \right]. \tag{78}$$

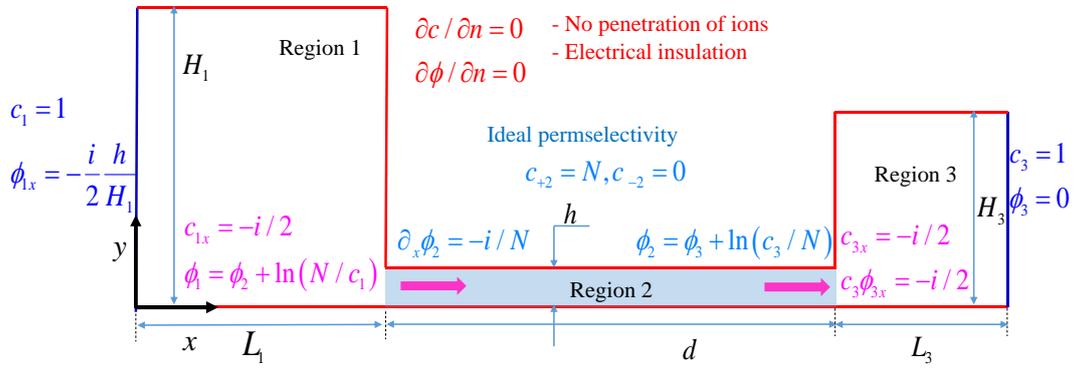

**Figure 1.** (Color online) Schematics describing the two-dimensional geometry of the three layers system consisting of a straight ideal permselective medium connecting two opposite asymmetric microchambers along with the boundary conditions for the LEN electrodiffusive problem.

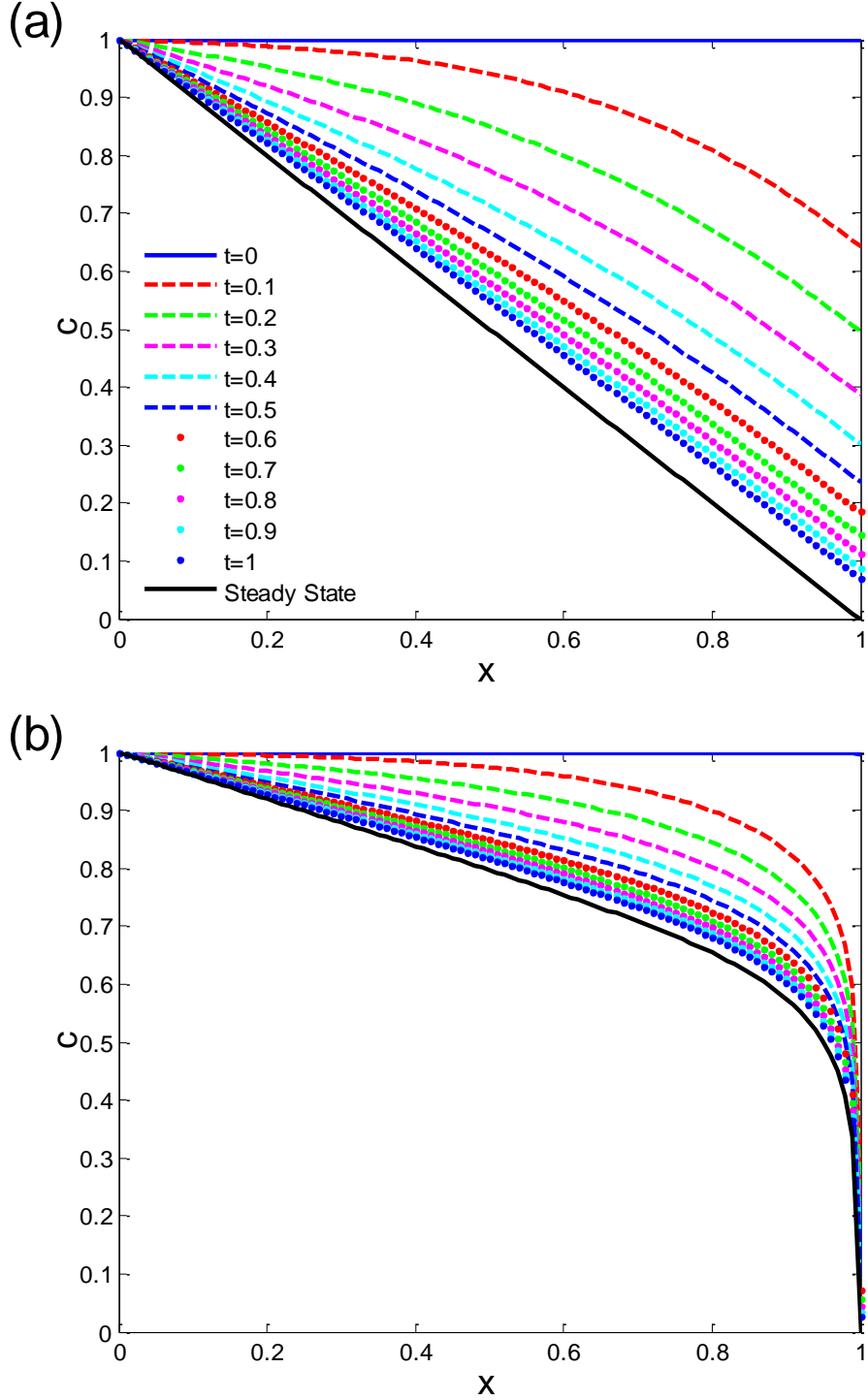

**Figure 2**. (Color online) Time-dependent concentration profiles in the depleted region (region 1) at the limiting current density for a (a) one-dimensional system $\left(i_{\lim}=2, h=H, L_1=1\right)$ and (b) a two-dimensional system $\left(I_{\lim}=0.322, H_1=0.4, h=10^{-3}, L_1=1\right)$.

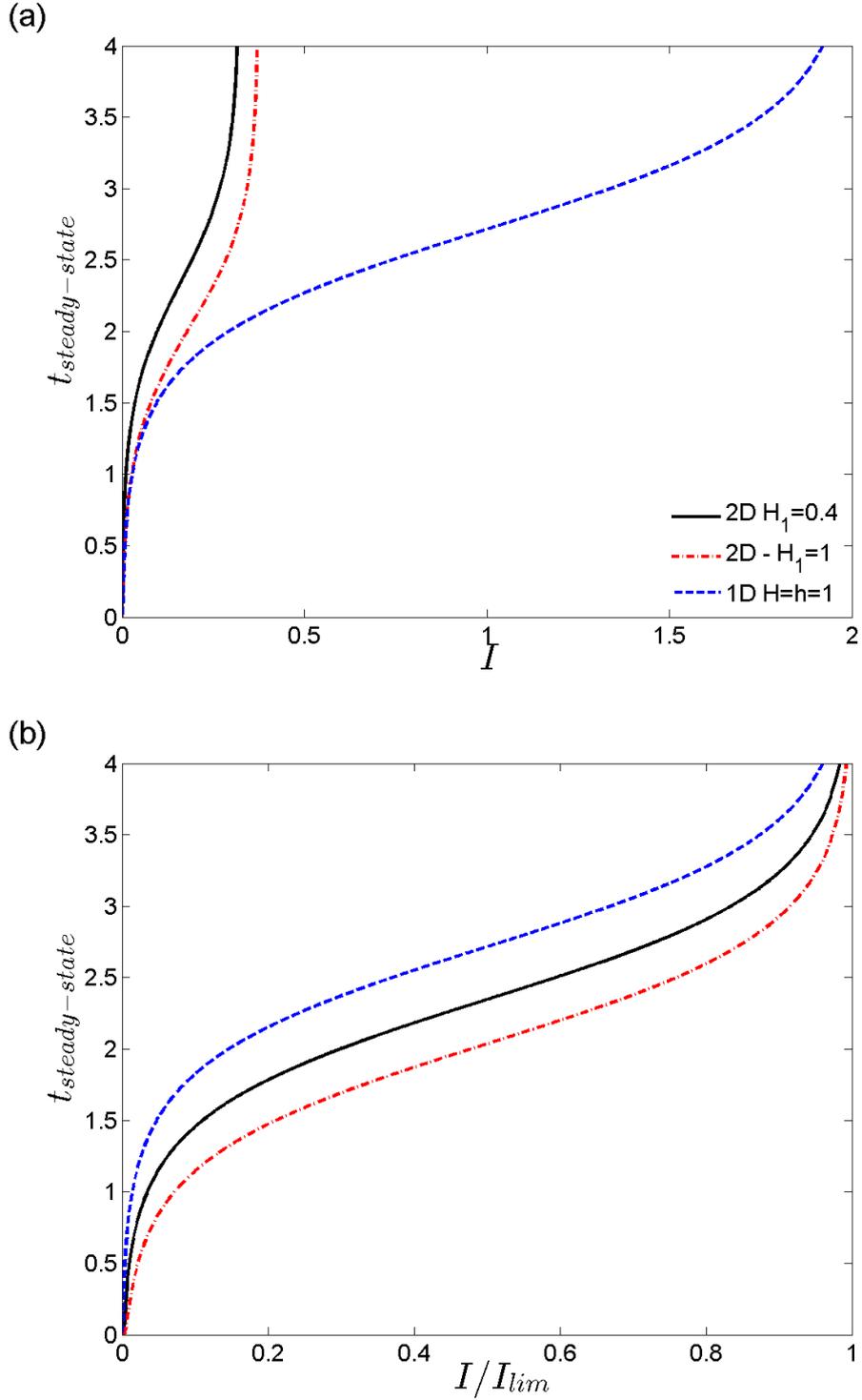

**Figure 3.** (Color online) The time to reach a quasi-steady state $(1.001c_s)$ as a function of the (a) applied current (b) applied current normalized by the limiting value of each configuration. The plots include the solution for one-dimensional and the two-dimensional cases of $(L_1 = 1, H_1 = 1)$ and $(L_1 = 1, H_1 = 0.4)$ with $h = 10^{-3}$, respectively.

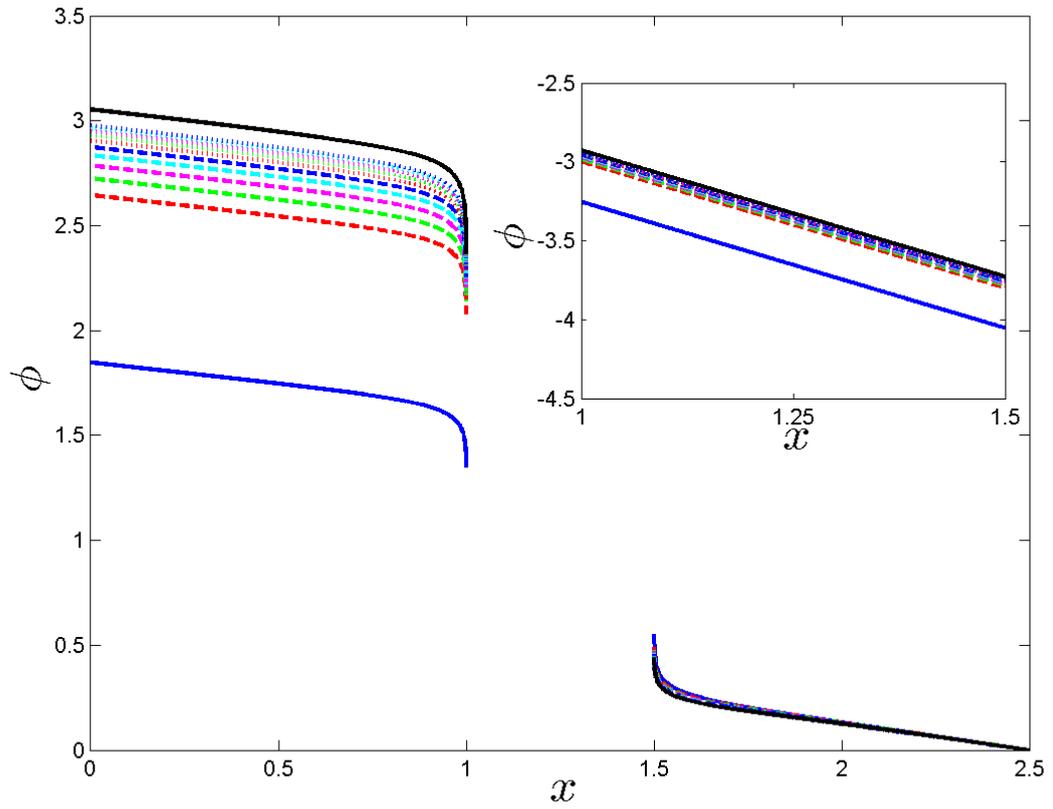

**Figure 4**. (Color online) Electric potential profile along the center line $(y=0)$ for the geometry $\left(L_1 = L_3 = 2d = 1, H_1 = 0.4, H_3 = 0.3, h = 10^{-3}, N = 10^2\right)$ at $I_{\lim}/2$. The inset shows the potential within the permselective medium. The legend is that shown in Figure 2.

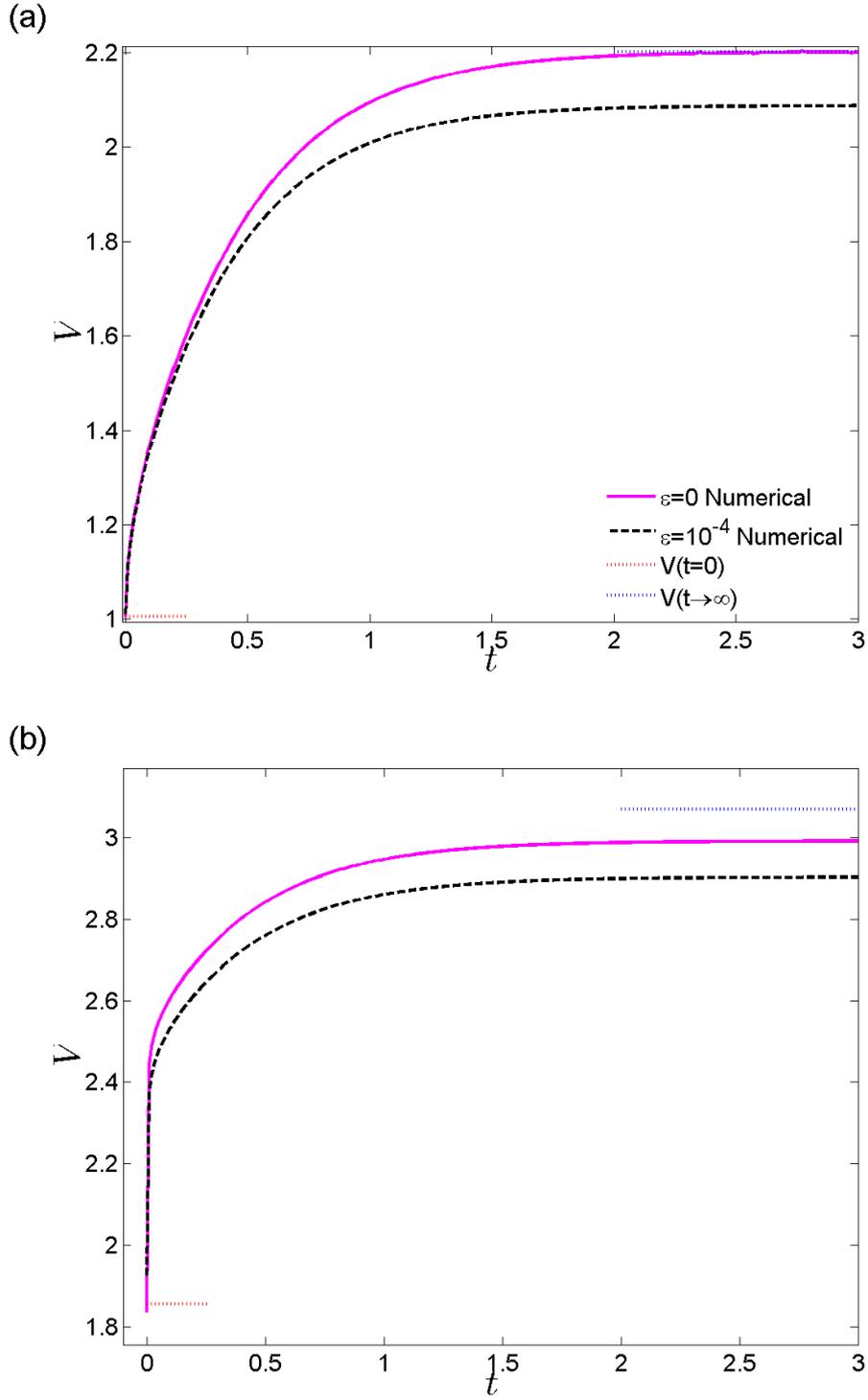

**Figure 5.** (Color online) *V-t* response for the (a) one-dimensional and (b) two-dimensional cases at their respective $0.5 \cdot I_{\text{lim}}$. The one-dimensional geometry is $(L_1 = L_3 = 1, d = 0.5, N = 100)$ while the two-dimensional geometric details are the same as in Figure 4.

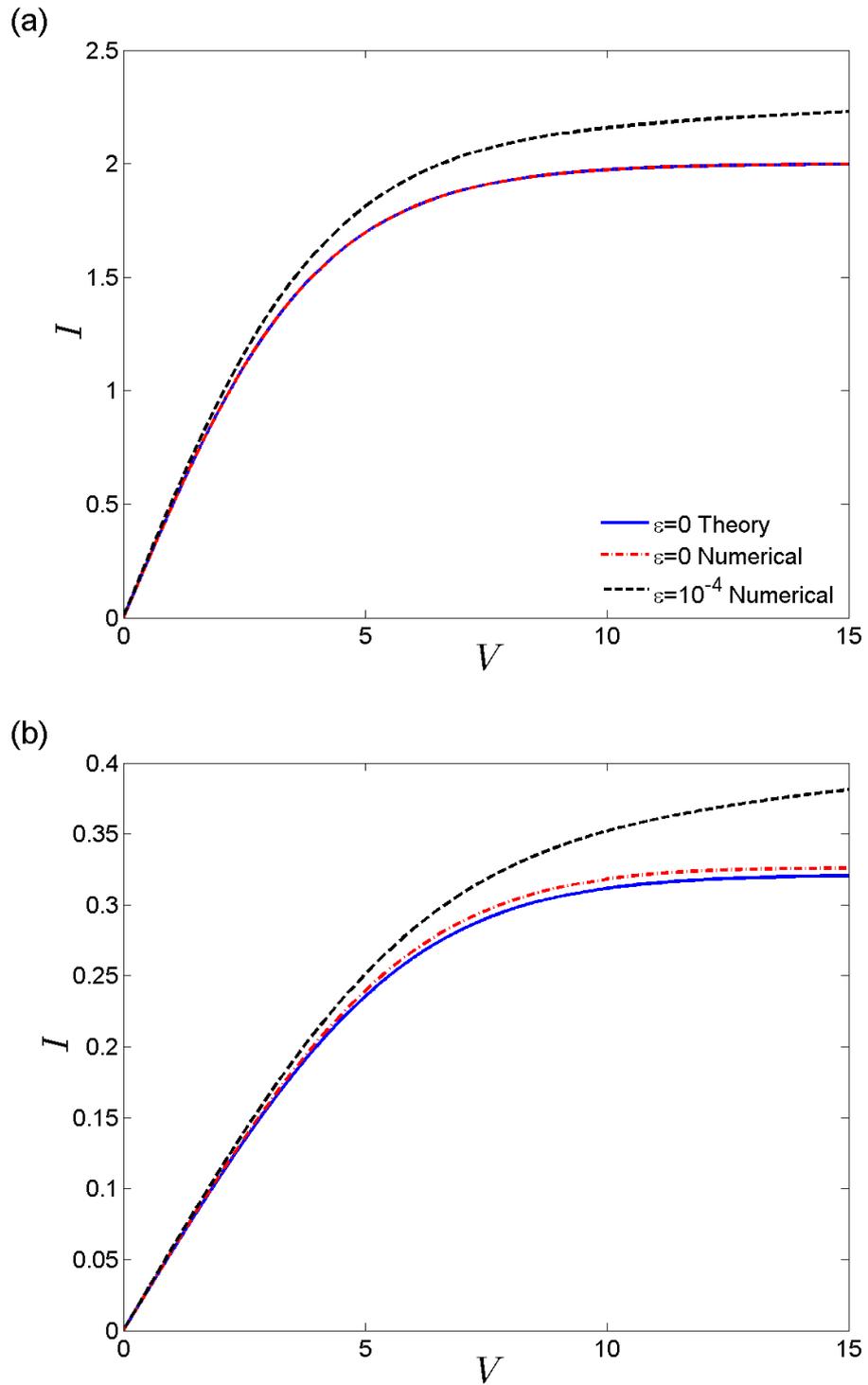

**Figure 6.** (Color online) Steady-state *I-V* curves computed for the LEN approximation and PNP equations for (a) one-dimensional and (b) two-dimensional cases. The geometries are the same as in Figure 4 and Figure 5.